\documentclass[twocolumn,showpacs,preprintnumbers,amsmath,amssymb,amsfonts]{revtex4}

\usepackage{graphicx}
\usepackage{dcolumn}
\usepackage{bm}
\usepackage{epstopdf}

\begin{document}

\title{Formation and local symmetry of Holstein polaron in $t$-$J$ model}

\author{Han Ma$^{1}$,  T. K. Lee$^{2}$,  Yan Chen$^{1,3}$}

\affiliation{$^{1}$Department of Physics, State Key Laboratory of Surface Physics and Laboratory of Advanced Materials, Fudan
University, Shanghai 200433, China\\
$^{2}$Institute of Physics, Academia Sinica, NanKang, Taipei 11529, Taiwan\\
$^{3}$Department of Physics and Center of Theoretical and Computational Physics, The University of Hong Kong, Pokfulam Road,
Hong Kong, China }

\date{\today}

\begin{abstract}

The formation and local symmetry of spin-lattice polaron has been investigated semiclassically in the planar
Holstein $t$-$J$-like models within the exact diagonalization method.
Due to the interplay of strong correlations and electron-lattice interaction,
the doped hole may either move freely or lead to the localized spin-lattice distortion and form a Holstein polaron.
The formation of polaron breaks the translational symmetry by suppression of
antiferromagnetic correlations and inducement of ferromagnetic correlations locally.
Moreover, the breaking of local rotational symmetry around the polaron has been shown.
The ground state is generically a parity singlet and the first excited state maybe a parity doublet.
Further consequences of the density of states spectra for comparison with future STM experiments are discussed.

\end{abstract}

\pacs{74.78.Fk, 71.10.Fd, 74.20.Fg}

\maketitle

Doping a Mott insulator has been regarded as the main physics in high $T_{c}$ cuprate superconductors~\cite{PALee}.
Single hole in the two-dimensional $t$-$J$ model may form a ferromagnetic (FM) spin polaron for sufficiently small
exchange coupling $J$~\cite{Nagaoka}. For finite $J$, the distortion of the antiferromagnetic (AFM) background
decays away from the hole and the competition between magnetic correlation
and kinetic energy may result in the ground state with a spin polaron structure.
This problem has been extensively investigated numerically~\cite{dagotto1,dagotto2,Roder,White,TKLee}.
To make a better comparison with angle-resolved photoemission spectroscopy (ARPES) experimental data, one need to include
the nearest-neighboring (NN) and next-nearest-neighboring (NNN) hopping $t'$ and $t''$ terms~\cite{shen1,shen2,supcond}.
On the other hand, the role of electron-phonon ($el$-ph) coupling has gained much interest recently. One reason is that the ARPES
data in doped metallic cuprates which showed the broadening of spectral lines at a certain momentum
revealed the band dispersion renormalized by $el$-ph
interaction~\cite{Shen2004,Shen2007}. In addition, the interaction also shifts the energy of the states.

The doped charge carriers in the presence of both strong electronic correlations
and electron-phonon interactions may lead to the formation of spin-lattice polaron.
In particular, the AFM exchange interaction allows for spin flips leading to
coherent hole motion at the bottom of the band and forms a spin polaron.
In the presence of strong
$el$-ph coupling, both the spin and lattice degrees of freedom become entangled
and the spin polaron may transform into a spin-lattice
polaron. The formation of this composite polaron may
affect both the spin and lattice degrees of freedom locally.
Recent ARPES experiments in undoped cuprates were interpreted in terms of
strong $el$-ph coupling giving rise to localized polaron~\cite{Shen2004,Shen2007}.
The possibility of self-localization of holes in lightly doped cuprates has been studied~\cite{polaron}.
It has been found that the
effect of $el$-ph interaction on spin polaron is strongly
enhanced as compared to polaron in uncorrelated systems~\cite{nagaosa1,Rosch,nagaosa2,Zhong,Fehske,Gunnarsson}.

Recent atomically resolved scanning tunneling microscopy (STM)
studies~\cite{STM1,STM2} on strongly underdoped cuprates revealed a
surprisingly pattern with the square symmetry of the lattice
broken on a local scale.
The origin of this broken local symmetry was attributed to the dopant impurity effect~\cite{Chen}.
The broken symmetry states were shown to appear in the case of a
hole confined to a cluster of sites centered at a impurity.
Meanwhile the introduction of $el$-ph interaction to the $t$-$J$ model may
stabilize the half-doped stripes~\cite{TKLee-stripe}.
In the presence of strong $el$-ph coupling, the
variations of hopping integral and spin-spin correlation
around the impurity may become more remarkable and the composite spin-lattice polaron may show up.

In this paper, we shall discuss the formation of spin-lattice polaron and
its relevance of local symmetry
by investigating the Holstein $t$-$J$-like models
with the exact diagonalization method.
 Due to the interplay of competing electronic correlations and $el$-ph interactions,
 the doped hole may either move through lattice freely or favor the composite spin-lattice polaron.
The formation of Holstein polaron breaks the translational symmetry by suppression of AFM correlations and
inducement of FM correlations locally.
Moreover, the breaking of local rotational symmetry around the polaron center has been shown.
The ground state is generically a parity singlet and the first excited state maybe a parity doublet.
Further consequences of the tunneling spectra for comparison with future STM experiments will be discussed.

The two-dimensional single-band Holstein $t$-$t'$-$J$ model
in the adiabatic limit is defined by the Hamiltonian,
\begin{eqnarray}
{\cal H} &=&  -t\sum_{\langle i,j \rangle \sigma} (c^{\dag}_{i\sigma} c_{j\sigma} + {\rm h.c.}) -t'\sum_{ \langle \langle i,j \rangle \rangle \sigma}
(c^{\dag}_{i\sigma} c_{j\sigma} + {\rm h.c.})
\nonumber \\
& & +
J\sum_{\langle i,j\rangle} \mbox{\boldmath $S_{i}\cdot S_{j}$}
 - g \sum_{i} u_{i}n^{h}_{i} + \frac{K}{2} \sum_{i} u^{2}_{i}.
\end{eqnarray}
where $c^{\dagger}_{i,\sigma}$ is an electron creation operator with spin $\sigma$ at site $i$ with a constraint of no double electron occupation due to strong electron correlations,
 $S_{i}$ is a spin operator for electron at site $i$, $\langle i,j \rangle$ and $\langle\langle i,j \rangle\rangle$ refer to NN and NNN sites $i$ and $j$. The first three terms in Eq. 1 represent the usual $t$-$t'$-$J$ model Hamiltonian. The fourth term denotes the $el$-ph interaction.
In the adiabatic limit, displacements $u_{i}$ have been treated classically and determined by $u_{i}= g/K \langle n^{h}_{i} \rangle $. The hole density operator $n^{h}_{i}$ is defined as $n^{h}_{i} = 1- n_{i}=1-\sum_{\sigma}c^{\dagger}_{i,\sigma} c_{i,\sigma}$. The last term is lattice elastic energy with elastic force constant $K$.
The energy spectrum and eigenstates can be obtained through exact diagonalization method.
The symmetry of the lowest energy state is sensitive to boundary condition and parameters. In our
calculations, we choose $t=K=1$, suitable for the
cuprates, and adopt the periodic boundary condition.

\begin{figure}[t]
\begin{center}
\includegraphics[height=3.1in,width=2.6in]{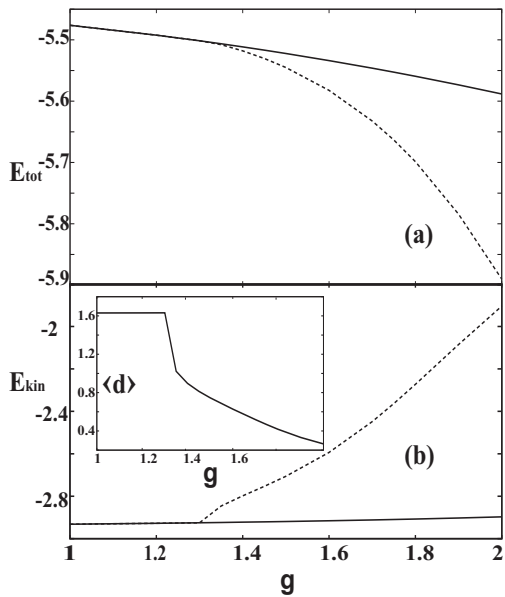}
\caption{The total energy (a) and kinetic energy (b) as a function of $el$-ph coupling constant $g$.
The solid line corresponds to the delocalized state while the dashed line denotes the localized Holstein polaron state.
The inset shows a measure of polaron size $\langle d \rangle$ as a function of $g$. We choose $t'=-0.1$ and $J=0.3$. }
\end{center}
\end{figure}

Employed with the  exact diagonalization method for a finite 16-site square cluster, the low-lying electronic
states have been calculated as a function of  $el$-ph coupling constant $g$.
Due to the competition between electronic correlations and electron-lattice interactions, the ground state
has certain limiting cases.
When $g$ is very small, the ground state must correspond to a delocalized state, that is to say,
doped hole may move through lattice freely so that the average occupation number of holes at each site is uniform.
The presence of strong $el$-lattice interaction may clearly favor the localized hole state, which is a sliding periodic polaron lattice. Since we focus on the local properties of polaron, we
use a very weak impurity to break the translational symmetry and pin down the sliding polaron lattice so that the doped single hole may stay around certain site.
At $g \gg 1$ limit, the doped hole tends to be localized and results in large lattice distortion
around that site. The corresponding ground state is a localized state with polaron formation.
The evolution of total energy as well as kinetic energy of the system as a function of $g$ is depicted in Fig. 1(a) and Fig. 1(b), respectively.
The critical $el$-ph coupling $g_{c} \sim 1.3$ can be straightforwardly determined by distinguishing these two distinct states.
Moreover we use the quantity $\langle d \rangle$ to measure the size of polaron qualitatively,
 $\langle d \rangle = \sum_i r_i \cdot n^h_i(r_i)$, where $r_i$ is the distance between site $i$ and polaron center.
It is obvious that the larger $\langle d \rangle$ value may correspond to a larger polaron size.
As shown in the inset of Fig. 1, as $g$ increases, doped holes tend to concentrate at the polaron center
so that the quantity $\langle d \rangle$ may be significantly suppressed.
It is worth to mention that the critical point $g_c$ is rather clearly revealed in the behaviors of $\langle d \rangle$ and kinetic energy than that of the total energy.
Comparing with the previous study on the dopant impurity effect~\cite{Chen}, the
strong variations of hopping integral and spin-spin correlation
around the polaron may lead to the formation of local lattice
distortion and the appearance of tightly bounded spin-lattice polaron due to strong
$el$-ph coupling~\cite{polaron}.

Next we examine the effect of NNN hopping integral $t'$. As we know, the $t'$ term plays an important role in understanding the superconducting
correlations in cuprate superconductors~\cite{Pavarini,TKLee1,TKLee2}. In particular,
the positive $t'$ case corresponds to the electron-doped system while the negative
$t'$ case corresponds to the hole-doped system.
In the following,
we systematically study the dependence of $g_c$ as a function of $t'$ for two distinct $t$-$J$ like models.
Due to the competing tendency between the polaron formation with strong $el$-ph interaction and the itinerant electrons with
large kinetic energy,  we naturally expect that the formation of localized polaron may require stronger $el$-ph interaction
to compensate larger kinetic energy $|t'|$ term.
For the Holstein $t$-$t'$-$J$ model, it is indeed the case as shown in Fig. 2(a) where $g_{c}$ goes up as $|t'|$ increases.
Another feature shown in Fig. 2(a) is the asymmetrical behavior between $t'<0$ and $t'>0$ regions.
An intuitive physical understanding of such
effect can be given as follows: for positive $t'$, the $t'$ and $t$
terms in kinetic energy match quite well, so that the positive $t'$ term
may enhance the kinetic energy more effective than the negative $t'$ term, hence
require stronger $el$-ph interaction $g_c$ to realize the localization of Holstein polaron.

\begin{figure}
\begin{center}
\includegraphics[height=1.6in,width=3.4in]{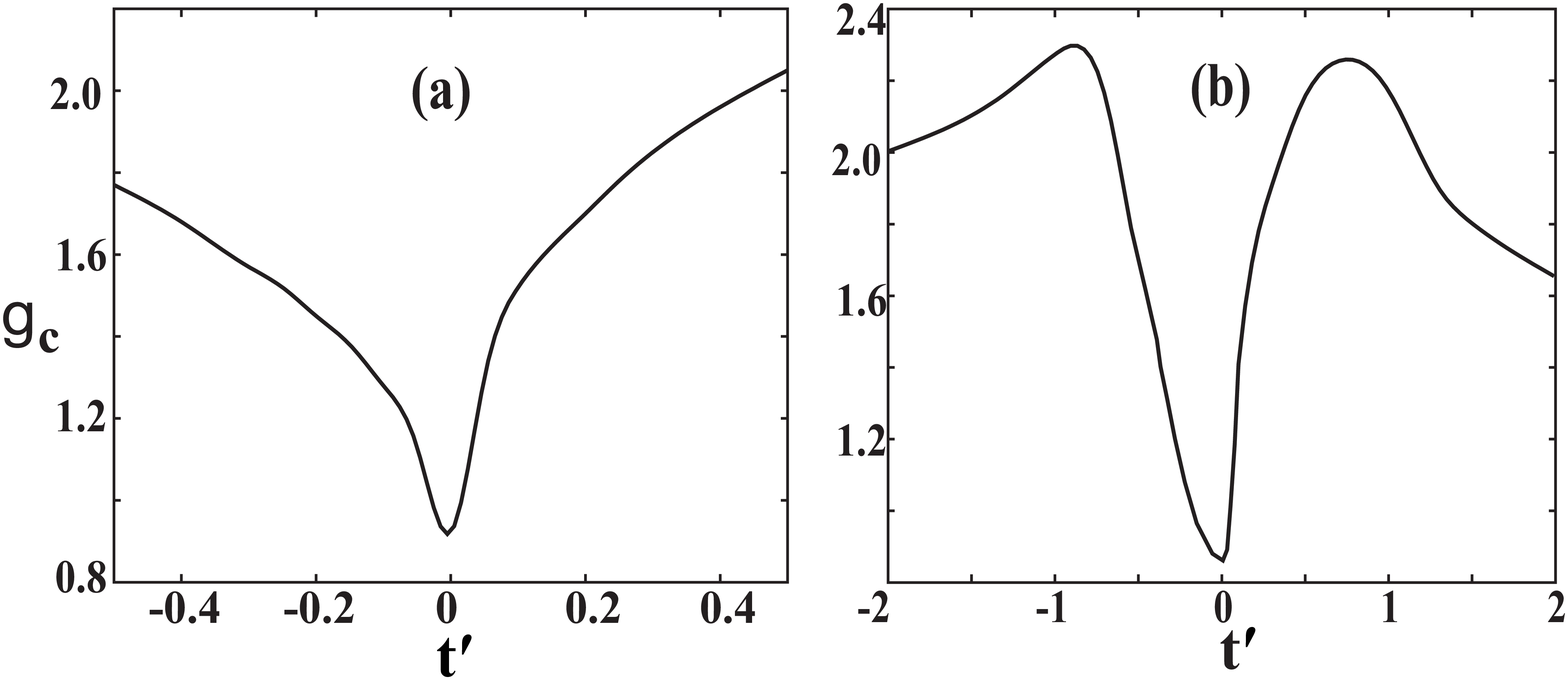}
\caption{The critical $el$-ph coupling $g_c$ as a function of $t'$ for Holstein $t$-$t'$-$J$ model (a) and $t$-$t'$-$J$-$J'$ model (b). We choose $J$=0.3 and $J'/J=(t'/t)^2$.}
\end{center}
\end{figure}

As suggested in recent studies, the $t$-$t'$-$J$-$J'$ model may be suitable for iron-based superconductors~\cite{QMSi,JPHu}.
In such systems, the appropriate range of $t'$ is much larger than that of cuprates and the NNN superexchange
coupling is expressed as $J'/J=(t'/t)^2$. In the case of weak $t'$ term, the results are similar to that of the Holstein $t$-$t'$-$J$ model.
The competing nature between $J'$ term and $J$ term may lead to the effect of geometrically frustration.
In such systems, the pairwise $J'$ interaction does not coincide
with the square lattice geometry, which may suppress the NN AFM correlation functions.
For stronger $t'$ term as well as $J'$ term, the suppression may become so strong that the local
FM correlation may emerge around the polaron center. And it may result in the enhancement of kinetic energy and the reduction of the critical $el$-ph
interaction $g_c$, which makes the formation of polaron easier. These relationships are clearly exhibited
in Fig. 2(b) where $g_{c}$ decreases as $t'$ becomes quite large.

Similar to the dopant impurity case~\cite{Chen},
the presence of a localized spin-lattice polaron may affect
not only the local charge and local spin distributions but
also the symmetry of the ground state wavefunction. In the following, we adopt the same
parity symmetry to characterize different quantum states~\cite{Chen}. In particular, we focus on
the reflection symmetries of a two-dimensional square lattice with
respect to $x$- and $y$-axes passing through the center of polaron
($P_x$ and $P_y$ respectively) and on the parity $P_xP_y$. Since
$[P_{x(y)},H]=0$, we may classify states according to the quantum
numbers of $P_x$, $P_y$. We denote the state with $(P_x=+1,P_y=+1)$
as state $(++)$, doubly degenerate state $(+1,-1)$,\,and  $(-1,+1)$
as states $(+-)$ and $(-+)$, and $(-1,-1)$ as state $(--)$.
As we know, in the absence of $el$-ph interactions,
for a 16-(4$\times$4) site cluster with periodic boundary condition,
the ground state of a single hole in the $t$-$t'$-$J$ model
has a four-fold symmetry for $t'<0$ , which can be represented by their
parity symmetry  $(++)$,$(+-)$,$(-+)$,$(--)$.

As clearly shown in Fig. 3, the ground states have 4-fold degeneracies for weak $el$-ph interaction $g < g_c \sim 1.3$,
For $g$ becomes slightly larger than $g_{c}$, such degeneracies are broken. We note that the four low-lying states with different reflection
symmetries are quite close in energy. The state with  $(++)$  symmetry always has the lowest energy, the doubly degenerated
states with parity symmetry $(+-)$ and $(-+)$ correspond to the first excited states, while the state with $(--)$ symmetry is the second excited state.
As $g$ exceeds  1.5, the state $(++)$ originally at much higher energy may drop down significantly and cross the the singly degenerated
$(--)$ state. For $g > 1.7$, both two lowest energy states have the $(++)$ symmetry.
According to our numerical results, the ground state of spin-lattice polaron is quite robust in the $(++)$ symmetry.

\begin{figure}
\begin{center}
\includegraphics[height=1.9in,width=3.3in]{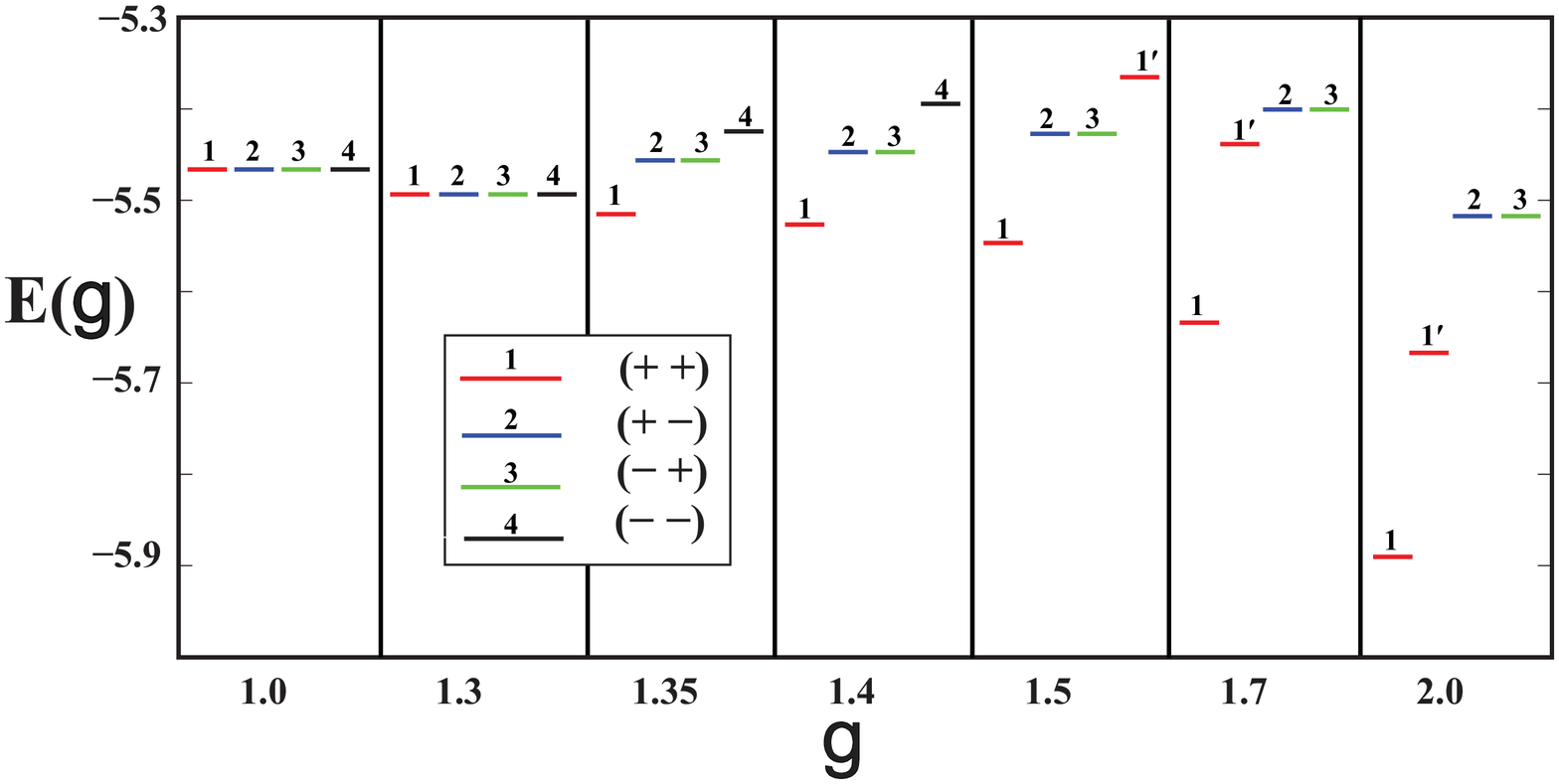}
\caption{(Color online) Evolution of four low-lying energy levels and their corresponding parity symmetries as a function of $el$-ph coupling
 $g$. We choose $t'=-0.1$ and $J=0.3$.}\label{fig:energylevel}
\end{center}
\end{figure}

To explore further the interplay between spin and charge degrees of freedom, we study the spatial distribution of
spin-spin correlation function and hopping integral around the polaron. As depicted in Table I, we
show the expectation value of spin-spin correlation function
$\langle S_i S_j \rangle$ as well as hopping integral $\langle
c_i^{\dagger}c_j \rangle$ for six distinct bonds illustrated in Fig. 4(a).
The interplay of strong correlation and $el$-ph interaction
may lead to hole localization and result in a remarkable hopping integral
and a weak FM spin-spin correlation function around the hole.
In particular, the AFM correlation along bond 1 is completely suppressed,
and a weak FM correlation emerges. The spin-spin correlation function
recovers quickly to the value -0.34 for farther bonds.
Meanwhile the hopping integral along bond 1 is much stronger than that of
the rest bonds and the hopping integral vanishes quickly close to the lattice boundary.
It is rather clear that the inducement of local FM correlations around the
polaron may help the hole moving around
more efficiently, then maximizing the local kinetic energy.
Hence the mutual cooperative effect between spin and lattice degrees of freedom has been clearly revealed.
For Holstein $t$-$J$ model in infinite dimensions, the interplay between the formation
of a lattice and magnetic polaron in the case of a single hole in the AFM
background has been studied before~\cite{Cappelluti}. It shows that the presence of
 AFM correlations favors the formation of the lattice polaron at lower values
 of the $el$-ph coupling. Our numerical calculations agree well with their results.

\begin{table}[htbp]
\centering
\caption{The spin-spin correlation functions and hopping integrals
 for various bonds in a 16-site cluster with periodic boundary condition at $t'=-0.1, J=0.3$ and
$g=2.0$. The bond indices are labeled in Fig. 4(a).
}
\begin{tabular}{c|c|c|c|c|c|c}
  \hline
  bond index & 1 & 2 & 3 & 4 & 5 & 6\\
  \hline
  $-\langle \textbf{S}_{i} \cdot \textbf{S}_{j} \rangle$ & -0.034 & 0.335 & 0.342 & 0.337 & 0.346 & 0.347\\
  $\langle c^{\dagger}_{i} c_{j} \rangle$ & 0.405 & 0.018 & 0.021 & 0.002 & 0.002 & 0.000\\
  \hline
\end{tabular}
\end{table}

 Furthermore, we investigate the parameter dependence of the spin-spin correlation function  for both
 $t$-$t'$-$J$ model and $t$-$t'$-$J$-$J'$ model.
 For these two models, the spin-spin correlation function on bond 1 and bond 2 shows quite different behaviors.
 For $t$-$t'$-$J$ model, there is always large AFM correlation on bond 2,
 while weak AFM or FM correlation shows up
 on bond 1. This strong suppression of AFM correlation locally around the center of polaron is due to the dramatic
 lattice distortion or the polaron formation.
 However, such strong suppression of AFM correlation on bond 2 may be
 significantly modified in $t$-$t'$-$J$-$J'$ model for large $t'$ case.
 In such case, the introduction of large $t'$ may  lead to the enhancement
 of  kinetic energy and  the strong frustration effect due to $J'$ term may greatly suppress the
 AFM correlation on bond 2. Meanwhile the size of the spin-lattice polaron may be enlarged.

\begin{figure}[b]
\begin{center}
\includegraphics[height=2.9in,width=3.1in]{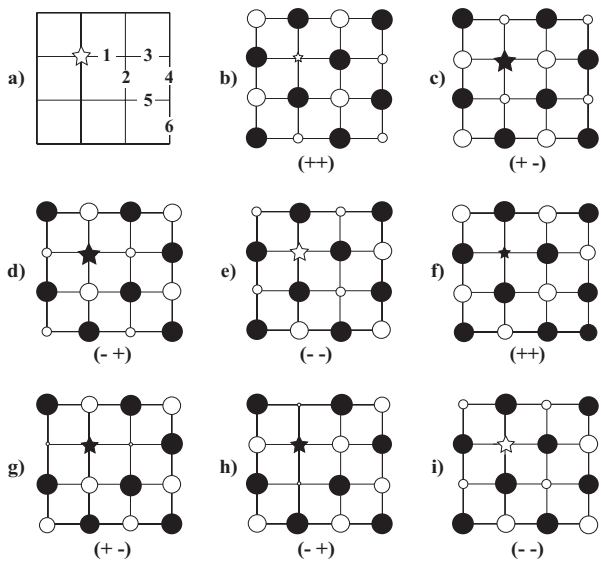}
\caption{The spatial distribution of  $S_{z}$ for $t$-$t'$-$J$ model. Delocalized states are shown in figure (b)-(e) for different parity symmetries. Localized states are shown in figure (f)-(i). The diameter of each circle is proportional to the value of local $|S_{z}|$. White (Black) circle represents up (down) spin. The star symbol labels the center of polaron.}\label{fig:sz}
\end{center}
\end{figure}

To understand the nature of states with different parity symmetries, the spatial distribution of $S_{z}$
has been calculated and illustrated in Fig. 4.
For delocalized states,  as shown in Fig. 4(b)-4(e), states with four distinct parity symmetry
 $(++)$, $(+-)$, $(-+)$, $(--)$ for $t$-$t'$-$J$ model display quite different patterns.
In Fig. 4(b), for state with $(++)$ symmetry, weak up spin appears at the polaron center while weak AFM
 correlation shows up between the polaron center and its four nearest neighbors.
 In Fig. 4(c)-(d), for states with $(+-)$ and $(-+)$ symmetry,
 the asymmetry between two neighboring sites of polaron center along $x$ and $y$ directions
  is clearly shown.
For localized states, the spatial distribution of $S_{z}$ patterns are shown in Fig. 4(f)-4(i), respectively,
corresponding to  different symmetry states.
For $(++)$ symmetry state, weak down spin appears at the polaron center while weak FM
 correlation shows up between the polaron center and its four nearest neighbors.
 As we know, the doped hole may concentrate at the polaron center where the weak FM correlation may emerge.
 It is obvious that this $(++)$ symmetry state is energetically favorable
 to the spin-lattice polaron formation.
 In Fig. 4(g)-(h), the asymmetry between two neighboring sites of polaron center along $x$ and $y$ directions
  becomes more significant than the delocalized states.
Of all four parity symmetry states, this $(++)$ symmetry state has the lowest energy.
In addition, we checked the spatial distribution of $S_z$ in $t$-$t'$-$J$-$J'$ model and similar results are obtained.
In both cases, the ground state of spin-lattice polaron prefers the $(++)$ parity symmetry.


In order to check the existence of spin-lattice polaron,
integrated differential conductance as a function of cutoff voltage is calculated to be compared with future STM experiments.
Following the STM tunneling theory~\cite{th}, we write the integrated current at $\textbf{r}$
up to a positive voltage $V$ as
\begin{eqnarray}
I(\textbf{r}, \omega)&\propto&\sum_{\sigma, m }
|\langle m | a^{\dagger}_{\textbf{r},\sigma} |\psi^{1h} \rangle |^2 \theta(\omega - E_{m} + E^{1h} ).
\end{eqnarray}
where $a^{\dagger}_{\textbf{r},\sigma}$ is the electron creation operator with
spin $\sigma$ at site $\textbf{r}$, $| m \rangle$ are eigenstates of the half-filled system with energy $E_{m}$, $\omega =eV$, and
$\theta$ is a step function. The $|\psi^{1h} \rangle$ denotes the single-hole eigenstate with energy $E^{1h}$.
In the following, we show the $I$-$\omega$ curve  on various sites and for different $el$-ph coupling $g$.
For convenience, we shift the origin of x-axis by  $\omega_0 = \omega + \omega_{ex}$ where $\omega_{ex} = E_0 - E^{1h}_{0}$ and $E_0$, $E^{1h}_{0}$ correspond to
 the ground state of half-filled system and single-hole system, respectively. Here we consider the contributions from low-lying energy states
 of the single-hole system.

As we know already, the doped hole tends to stay around the polaron center and the hole density at the polaron center
is more significant than that of on other sites by increasing the $el$-ph interaction $g$.
Thus we expect the integrated current at the polaron center would become larger when $g$ goes up.
This result is clearly shown in the left panel of Fig. 5.
In the case of impurity doped cuprates, the conductance pattern is anisotropic as the tip of a tunneling
microscope scans above  the Cu-O-Cu bonds along the $x$($y$)-axes. This anisotropy is quite pronounced at voltage around $\omega \sim J$
In the present case, due to the formation of spin-lattice polaron, the ground state of one-hole system has (++)  symmetry, the first excited state
corresponds to the doubly degenerated states $(+-)$ and $(-+)$. The presence of quadrupole interaction of two
single-hole states or by other couplings may not change the symmetry of ground state but may destroy the two-fold degeneracy of first excited state.
We consider the system to be in one of the two degenerate states, say in the state of $(+-)$.
As depicted in the right panel of Fig. 5, below certain cutoff-energy ($\sim 0.11 J$), the integrated current shows four-fold rotational symmetry for the
state with $(++)$ symmetry. For the higher cutoff-energy, the first excited state with $(+-)$ symmetry may play important role in the local symmetry breaking
for the signals along $x$ and $y$- directions.

\begin{figure}
\begin{center}
\includegraphics[height=1.4in,width=3.3in]{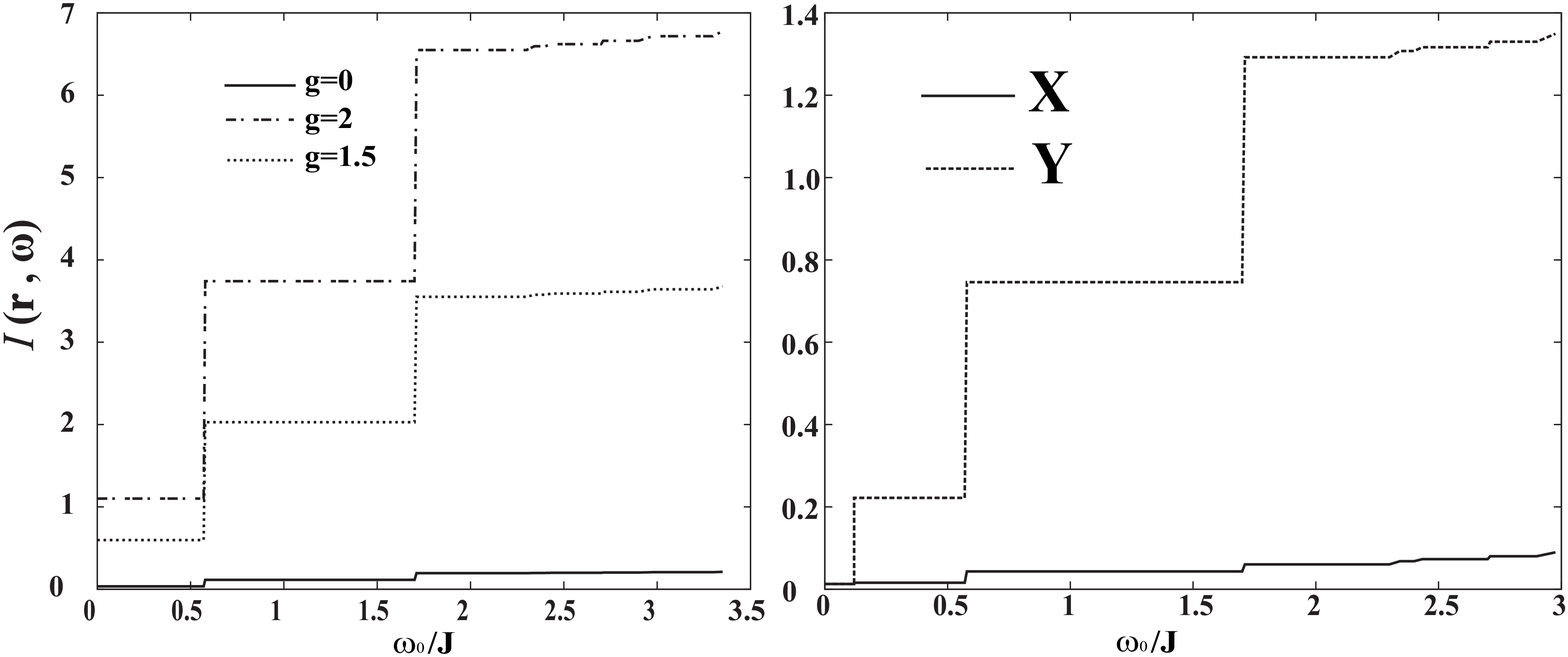}
\caption{The local integrated current up to voltage $V$ for different sites at various
$e$-ph coupling constant $g$ as a function of $\omega$. The left panel shows the results for signal at the
center of polaron for different $g$ values. The right panel depicts the results for signal at the two NN sites respectively along $x$ and $y$ directions.}
\end{center}
\end{figure}\label{fig5}


Our numerical results can be qualitatively understood in terms of the renormalized mean-field method for $t$-$J$ model.
Due to the presence of spatial inhomogeneous polaron, the renormalized factor $g_t$ and $g_s$ become site-dependent.
According to previous study~\cite{QHWang}, the renormalized factor of kinetic energy on the bond connecting two sites $i$ and $j$ can
be expressed as $g_t^{ij} \sim \sqrt{\delta_i \delta_j}$.
As we know, in the case of strong $el$-ph interaction, the hole density is highly localized at the center of Holstein polaron.
It is obvious that the bonds with appreciable hopping integral show up only around the polaron and may result in the
suppression of AFM correlation locally.
For the delocalized states, $g_t^{ij}$ is simply a constant and so is the AFM correlation function.
The above qualitative analysis agrees reasonable well with our
numerical results shown in Fig. 3 and Table.I.

In summary, we study both the formation and local symmetry of spin-lattice polaron semiclassically in the planar
Holstein $t$-$J$-like models within the exact diagonalization method. Due to the interplay of competing interactions among electronic correlations and
$el$-phonon interactions, the doped hole may either move freely or
lead to the localized spin-lattice distortion and form a Holstein polaron.
Since the formation of polaron breaks the translational symmetry, we use the parity symmetry with respect to the polaron center to characterize the localized states.
The presence of spin-lattice polaron may suppress the AFM correlations and induce the FM correlations locally around the polaron.
This effect may further stabilize  the spin-lattice polaron. Moreover, this affect may lead to a strong localized state with $(++)$ parity symmetry as ground state.
Moreover, the breaking of local rotational symmetry around the polaron has been shown for certain voltage cutoff.
The present investigation on the polaron formation and local symmetry may provide useful information
for future STM experimental tests.

This work was supported by the National Natural Science Foundation
of China (Grant Nos. 11074043 and 11274069) and the State Key Programs of China
(Grant Nos. 2009CB929204 and 2012CB921604) and Shanghai Municipal Government,
the National Science Council in Taiwan with Grant No. 98-2112-M-001-017-MY3, the RGC
grants in HKSAR, Fudan's Undergraduate Research Opportunities Program and
National Science Fund for Talent Training in Basic Science.


\begin{thebibliography}{3}

\bibitem{PALee} P. A. Lee, N. Nagaosa, and X. G. Wen, Rev. Mod. Phys. \textbf{78}, 17 (2006).


\bibitem{Nagaoka} Y. Nagaoka, Phys. Rev. \textbf{147}, 392 (1966).



\bibitem{dagotto1} E. Dagotto, A. Moreo, T. Barnes, Phys. Rev. B \textbf{40}, 6721 (1989).

\bibitem{dagotto2} E. Dagotto, Rev. Mod. Phys. \textbf{66}, 763 (1994).

\bibitem{White} S. R. White and I. Affleck, Phys. Rev. B \textbf{64}, 024411 (2001).

\bibitem{TKLee} T. K. Lee, C. M. Ho, and N. Nagaosa, Phys. Rev. Lett. \textbf{90}, 067001 (2003).



\bibitem{Roder} H. R$\ddot{o}$der, H. Fehske, and H. Buttner, Phys. Rev. B \textbf{47}, 12420 (1993).

\bibitem{shen1} A. Danmascelli, Z.-X. Shen, and Z. Hussain, Rev. Mod. Phys. \textbf{75}, 473 (2003).

\bibitem{shen2} C. Kim, P. J. White, Z.-X. Shen, T. Tohyama, Y. Shibata, S. Maekawa, B. O. Wells, Y. J. Kim, R. J. Birgeneau and M. A. Kastner, Phys. Rev. Lett. \textbf{80}, 4245 (1998).

\bibitem{supcond} T. Tohyama and S. Maekawa, Supercond. Sci. Technol. \textbf{13}, R17 (2000).

\bibitem{Shen2004} K. M. Shen, F. Ronning, D. H. Lu, W. S. Lee, N. J. C. Ingle, W.
Meevasana, F. Baumberger, A. Damascelli, N. P. Armitage, L. L.
Miller, Y. Kohsaka, M. Azuma, M. Takano, H. Takagi, and Z.-X.
Shen, Phys. Rev. Lett. \textbf{93}, 267002 (2004).

\bibitem{Shen2007} K. M. Shen, F. Ronning, W. Meevasana, D. H. Lu, N. J. C. Ingle,
F. Baumberger, W. S. Lee, L. L. Miller, Y. Kohsaka, M. Azuma, M. Takano, H. Takagi, and Z.-X. Shen,
Phys. Rev. B \textbf{75}, 075115 (2007).

\bibitem{polaron} P. Prelov$\breve{s}$ek, R. Zeyher, and P. Horsch, Phys. Rev. Lett. \textbf{96}, 086402 (2006).

\bibitem{nagaosa1} A.S. Mishchenko and N. Nagaosa, Phys. Rev. Lett. \textbf{93}, 036402 (2004).

\bibitem{Rosch} O. R\"osch, O. Gunnarsson, X. J. Zhou, T. Yoshida, T. Sasagawa, A. Fujimori, Z. Hussain, Z.-X. Shen, and S.
    Uchida, Phys. Rev. Lett. \textbf{95}, 227002 (2005).


\bibitem{nagaosa2} G.De Filippis, V. Cataudella, A.S. Mishchenko and N. Nagaosa, Phys. Rev. Lett. \textbf{99}, 146405 (2007).

\bibitem{Zhong} J. Zhong and H. B. Sch\"uttler, Phys. Rev. Lett. \textbf{69}, 1600 (1992).

\bibitem{Fehske} H. Fehske, H. R\"oder, G.Wellein, and A. Mistriotis, Phys. Rev. B \textbf{51}, 16582 (1995).


\bibitem{Gunnarsson} O. Gunnarsson and O. R\"osch, Phys. Rev. B \textbf{73}, 174521 (2006).


\bibitem{STM1} T. Hanaguri \textit{et al.}, Nature \textbf{430}, 1001 (2004).

\bibitem{STM2} Y. Kohsaka, C. Taylor, K. Fujita, A. Schmidt, C. Lupien, T. Hanaguri, M. Azuma,
M. Takano, H. Eisaki, H. Takagi,. Uchida, and J. C. Davis, Science \textbf{315}, 1380 (2007).



\bibitem{Chen} Y. Chen, T.M. Rice and F.C. Zhang, Phys. Rev. Lett. \textbf{97}, 237004 (2006).

\bibitem{TKLee-stripe} C.-P. Chou and T. K. Lee, Phys. Rev. B \textbf{81}, 060503(R)(2010); C.-P. Chou and T. K. Lee, Phys. Rev. B \textbf{85}, 104511(2012).

\bibitem{Pavarini} E. Pavarini, I. Dasgupta, T. Saha-Dasgupta, O. Jepsen, and O. K. Andersen, Phys. Rev. Lett. \textbf{87}, 047003 (2001).

\bibitem{TKLee1} W. C. Lee, T. K. Lee, C. M. Ho, and P.W. Leung, Phys. Rev. Lett. \textbf{91}, 057001 (2003).

\bibitem{TKLee2} C. T. Shih, T. K. Lee, R. Eder, C.-Y. Mou, and Y. C. Chen, Phys. Rev. Lett. \textbf{92}, 227002 (2004).



\bibitem{QMSi} Q. Si and E. Abrahams, Phys. Rev. Lett. \textbf{101}, 076401 (2008).

\bibitem{JPHu} C. Fang, H. Yao, W.F. Tsai, J.P. Hu, and S.A. Kivelson, Phys. Rev. B \textbf{77}, 224509 (2008).


\bibitem{Cappelluti} E. Cappelluti and S. Ciuchi, Phys. Rev. B \textbf{66}, 165102 (2002).


\bibitem{th} J. Tersoff and D.R. Hamann, Phys. Rev. B \textbf{31}, 805 (1985).


\bibitem{QHWang} Q. H. Wang, Z. D. Wang, Y. Chen, and F. C. Zhang, Phys. Rev. B \textbf{73}, 092507 (2006).


\end{thebibliography}
\end{document}